\documentclass[a4paper]{jpconf}
\usepackage{graphicx}
\usepackage{subeqnarray}
\input{epsf}

\begin{document}

\title{Formation of molecular hydrogen on analogues of
interstellar dust grains: experiments and modeling}

\author{Gianfranco Vidali$^{1}$, Joe Roser$^{1}$,
Giulio Manic\'o$^{2}$, Valerio Pirronello$^{2}$,
Hagai B. Perets$^{3}$,
and Ofer Biham$^{3}$}

\address{$^{1}$ Syracuse University, 201 Physics Bldg.,
                Syracuse, NY 13244-1130, USA}
\address{$^{2}$ Dipartimento di Metodologie Fisiche e Chimiche per l'Ingegneria,
                Universita' di Catania, 95125 Catania, Sicily, Italy}
\address{$^{3}$ Racah Institute of Physics, The Hebrew University,
                Jerusalem 91904, Israel}

\ead{gvidali@syr.edu}

\begin{abstract}
Molecular hydrogen has an important role in the early stages of star
formation as well as in the production of many other molecules that have
been detected in the interstellar medium. In this review we show that it
is now possible to study the formation of molecular hydrogen in simulated
astrophysical environments. Since the formation of molecular hydrogen is
believed to take place on dust grains, we show that surface science
techniques such as thermal desorption and time-of-flight can be used to
measure the recombination efficiency, the kinetics of reaction and the
dynamics of desorption. The analysis of the experimental results using
rate equations gives useful insight on the mechanisms of reaction and
yields values of parameters that are used in theoretical models of
interstellar cloud chemistry.
\end{abstract}

\section{Introduction}

The formation of molecular hydrogen in interstellar space is an
important problem in astrophysics and astrochemistry. Molecular
hydrogen is not only the most abundant molecules in the Universe,
but it contributes to the initial gravitational collapse of a cloud
by re-radiating in the infrared energy generated by the collapse and
intervenes, either in its charged or neutral form, in virtually all
reaction schemes leading to the formation of more complex molecules
in the interstellar medium (ISM). The reasons why there has always
been much interest in its formation is that molecular hydrogen needs
to be continuously generated in space, since ultraviolet photons,
cosmic rays, shocks and chemical reactions are the main agents
contributing to its destruction, and that its formation in the
gas phase is very inefficient. Although there are various routes
to make molecular hydrogen, formation rates of reactions in the
gas phase cannot produce it in enough quantities to explain its
presence given the destruction rates of the processes mentioned above
\cite{Herbst2001}.
In the Sec. 2 we present a brief survey of
the main developments in the studies of H$_2$
formation on interstellar dust grains over the past
forty years or so. The early theories are reviewed
in light of the constraints imposed by observations.
The main results of laboratory experiments over the
past decade and their implications are described.
In Sec. 3 we review the reaction mechanisms on surfaces that
are relevant to molecular hydrogen formation.
The experiments performed in our laboratory are introduced in
Sec. 4 and the results are presented in Sec. 5.
These results are analyzed in Sec. 6.
The implications and open problems are discussed in Sec. 7,
followed by a summary in Sec. 8.

\section{Formation of H$_2$ on interstellar dust grains}

\subsection{Review of early theories}

Gould and Salpeter
\cite{Gould1963}
showed that H$_2$
formation in the gas phase is not feasible and
proposed that interstellar dust grains can act as catalysts in the formation
of molecular hydrogen.
In subsequent work,
Hollenbach et al.
showed that dust grains
can successfully catalyze molecular hydrogen production
at a rate which is
compatible with its observed abundance
\cite{Hollenbach1970,Hollenbach1971a,Hollenbach1971b}.
The rate coefficient of molecular hydrogen formation
$R$ (cm$^3$ sec$^{-1}$)
is related to
the total number density of hydrogen atoms
(both in atomic and molecular forms)
$n=n_{\rm H} + 2 n_{\rm H_2}$
and the photodissociation rate
$\beta = 5 \times 10^{-10}$
(sec$^{-1}$) by:
$R n_{\rm H} n = 0.11 \beta n_{\rm H_2}$
\cite{Duley1984}.
Jura calculated, based on an analysis of Copernicus observations in the local
diffuse cloud medium, that the formation rate coefficient on grains is
$10^{-17} < R < 3 \times 10^{-17}$ (cm$^3$ sec$^{-1}$)
\cite{Jura1975}.
Since then,
$R$
has been
measured in several different ISM environments.
Remarkably, these
analyses give a value of $R$
very similar to the one estimated by
Jura for local diffuse clouds.
Recently,
Gry et al.
\cite{Gry2002}
have looked at the formation rate of H$_2$ in the diffuse ISM,
using FUSE data,
while Habart et al.
\cite{Habart2004}
examined this rate
in photodissociation regions, using SWS-ISO data.
The reader is referred to these papers for a discussion
of the mechanisms of formation of H$_2$ in these conditions.

To carry out the calculations, Hollenbach, Werner and Salpeter
\cite{Hollenbach1970,Hollenbach1971a,Hollenbach1971b}
made a number of assumptions, of which a couple turned out to
be problematic: first, the calculation was done on grains coated
with crystalline ice, since this was the understanding at that
time, although later it was shown that most of ice in space is in
the amorphous form and that the grains are bare in the diffuse cloud medium.
Second, it was assumed that within the grain, hydrogen atoms moved from
site to site by tunneling. Specifically, in each small crystal of ice,
diffusion in the spatially periodic potential was by tunneling and very fast;
Nonetheless, to prevent evaporation of atoms before
they met each other, it was necessary to set the condition that the
surface has enhanced binding sites beyond the weak ones. In any case, to
make hydrogen formation likely over a range of temperature in the diffuse
ISM, the H atoms cannot experience strong chemisorption forces, or
otherwise there would be little migration of atoms out of the
chemisorption sites.
Under these assumptions, they arrived at the following expression
for the recombination rate
$R_{\rm H_2}$ (cm$^{-3}$ sec$^{-1}$),
namely the number of
molecules formed per unit time and unit
volume:

\begin{equation}
R_{\rm H_2} = \frac{1}{2} n_{\rm H} v_{\rm H}
                          \sigma \xi \eta n_{\rm g}.
\label{eq:Production3}
\end{equation}

\noindent
In this formula,
$n_{\rm H}$
is the number density of hydrogen atoms,
$v_{\rm H}$ is their speed,
$\sigma$ is the cross section of the grain,
$\xi$ is the sticking coefficient,
$\eta$ is the probability of bond formation once
two atoms encounter each other
and $n_{\rm g}$
is the number density of dust grains in the ISM.
Assuming that the average sticking coefficient is
$\xi = 0.3$,
a result of their calculation of the interaction of a hydrogen
atom with an ice surface, they obtained that they could recover
the observed rate coefficient
$R$
of H$_2$ production
if the probability of recombination of hydrogen atoms when on
the surface was $\eta=1$.

A dissenting voice was that of Smoluchowski
\cite{Smoluchowski1981,Smoluchowski1983},
who considered the problem of hydrogen atoms
diffusing on an amorphous ice surface.
Since tunneling is very sensitive, on an atomic scale,
to the details of the environment through which a hydrogen atom diffuses,
it is not surprising that in an amorphous medium the mobility is greatly reduced.
Unfortunately, his calculations gave the result that the mobility of hydrogen
atoms on an ice surface would be much too small to yield recombination events
with the rate necessary to match observations.

An example of an alternative formation route for H$_2$ is the
chemistry induced by charged particle bombardment of interstellar
ices. Brown et al. conducted experiments where they sent MeV ions on
water ice and measured the number of H$_2$O, H$_2$ and O$_2$
molecules, that were sputtered, or ejected in the gas-phase, per
each impinging ion \cite{Brown1982}. Using these experimental
results, Avera and Pirronello evaluated theoretically the production
rate of molecular hydrogen per unit volume and time in dense
interstellar clouds due to the bombardment of the icy mantles on
grains by cosmic rays \cite{Pirronello1988,Averna1991}. The result
they obtained was orders of magnitude higher than Smoluchowski's
evaluation but much smaller than that of Hollenbach et al.
\cite{Hollenbach1971b}.

Experiments on the formation of molecular hydrogen on surfaces were done
since the work of Hollenbach and Salpeter, but these studies were carried
out under conditions far enough from ISM conditions that they were of
little use in elucidating the actual processes on dust grains.
These studies are reviewed in
Ref.
\cite{Pirronello2000}.

\subsection{Constraints on H$_2$ formation on dust grains}

Consider the formation of molecular hydrogen in typical conditions
found in diffuse or dense clouds.
The gas-phase hydrogen atoms have kinetic
energies below a few hundred K and grain temperature is low
($<20$ K).
Therefore, strongly
bound (chemisorbed) atoms will be virtually immobile.
Diffusion energy barriers,
$E_{\rm H}^{\rm diff}$,
are rather sensitive to the morphology of the surface.
Roughly speaking,
for weak adsorption sites
$E_{\rm H}^{\rm diff} \sim E_{\rm H}^{\rm des}/5$,
where
$E_{\rm H}^{\rm des}$,
is the binding energy of the atom.
Strong adsorption sites have proportionally smaller diffusion
energy barriers, namely
$E_{\rm H}^{\rm diff} \sim E_{\rm H}^{\rm des}/20$.
A typical grain of $0.1$ $\mu$m radius,
has an area $A \sim 10^{-9}$ cm$^2$, assuming that it is
spherical
(and considerably larger area for a rough surface).
The number of
adsorption sites
on the grain
is roughly $10^{15} A$ sites.
There are a few timescales to consider:
the time between arrivals, the residence time on
the surface, and time between the
arrivals of UV photons that can produce
fluctuations in the temperature of the grain.
The size of the grain, the porosity and the
distribution of binding sites in energy
are also important parameters. We now examine
how these timescales are inter-related
and what conditions we need to impose on the other
parameters in order to have a
reaction between two hydrogen atoms.

The arrival rate of H atoms on a grain is given by $n_{\rm H} v_{\rm
H} \sigma$. A fraction $\xi$ of the impinging atoms stick to the
surface. For $n_{\rm H} = 100$ (atoms/cm$^3$) and $v_{\rm H} \sim
10^5$ (cm/sec), assuming gas temperature of about 100 K, one obtains
that the collision rate of H atoms with a typical grain is $\sim 1$
particle per 300 seconds. The residence time is $t_{\rm H} =
\nu^{-1} \exp(E_{\rm H}^{\rm des}/k_B T)$, where $\nu$ is the
vibrational frequency (typically $\sim 10^{12}$ - $10^{13}$
sec$^{-1}$). $E_{H}^{\rm des}$ is the binding energy of the atom to
an adsorption site on the surface, namely the activation energy
barrier for desorption of the atom. Therefore, for $E_{\rm H}^{\rm
des}/k_B \sim 500$ K, and surface temperature of $T \sim 10$ K, the
residence time is $10^8 - 10^9$ (sec), but at higher temperatures
around $T \sim 15$ K, $t_{\rm H}$ is reduced to the range of 30 to
300 (sec), which is comparable with the time between arrivals. The
temperature of the grain is influenced by the ultraviolet (UV)
photon flux. The time-scale for absorption of a photon is $t_{\rm
photon} = (\sigma Q \Phi)^{-1}$ \cite{Pirronello2004}, where
$\sigma$ is the geometric cross section of the grain, $Q$ is the
absorption efficiency and $\Phi$ is the UV flux. In diffuse clouds,
the average UV flux is $2.6 \times 10^{-3}$ (erg cm$^{-2}$
sec$^{-1}$) \cite{Draine1978}. For large grains, the interval
between absorption of photons is short (tens of seconds) and the
grain will keep a stable temperature in the 15-20 K range. For
smaller grains, the temperature fluctuations are in the range of 10
to 20 K for grain diameters $a \sim 0.01 \mu$m, where $t_{\rm
photon} \sim 750$ (sec), and in the range of 15 to 20 K for $a \sim
0.02 \mu$m, where $t_{\rm photon} \sim 100$ (sec) \cite{Draine2003}.
In these cases and for even smaller grains, the arrival rate of
particles is crucial since we need $\xi dN/dt > 1/t_{\rm photon}$ in
order to maintain high efficiency of H$_2$ formation. It is clear
that in this model the presence of H atoms on the surface is
extraordinarily sensitive to the temperature of the grain, which in
turn is related to its size, and that there is a rather narrow
window when reactions can occur, as Hollenbach and Salpeter already
pointed out \cite{Hollenbach1971a}.

The sensitivity to the conditions on the grains may be reduced by
the fact that the surface of the grain may be porous enough that
even if there is spontaneous desorption of H from a site there is a
good chance that the H atom will strike another side of the grain
and remain stuck to it. Also, there may be some density of sites
with sufficient binding energy to confine H atoms for a long time.
The distribution of sites might influence the formation of H$_2$
quite a bit. Furthermore, the broad grain-size distribution,
namely by the fact that there are many more small grains enhances
the total surface area. The resulting increase in the collision rate
between H atoms and dust grains may partly compensate for the low
efficiency of H$_2$ formation.

\subsection{Recent experiments and their implications}

In the late 1990's, about thirty years after the work of
Salpeter and collaborators,
our group embarked on a program to measure how efficient
the process of formation of
molecular hydrogen on dust grains is, and whether, by recreating some of these
processes in the laboratory, it is possible to learn about both the physical
processes at play and the suitability of proposed dust grain analogues as
replacements for the yet poorly known actual star dust material.

The first dust grain analogue that was used to test whether molecular
hydrogen could be produced via processes described by the theory of
Salpeter and collaborators was a polished polycrystalline olivine stone,
a silicate.
Silicates in space are mostly olivines, with a composition towards a
magnesium-rich
mineral, (Mg$_x$Fe$_{1-x}$)$_2$SiO$_4$
and $x$ ranging from 1 for fayalite to 0 for fosterite
\cite{Colangeli2003}.
The reason to choose olivine as a sample to study is that
silicates are thought to be abundant in diffuse clouds, where the high
destruction rate of molecular hydrogen makes it crucial that there exists an
effective mechanism of formation.
The results of our experiments, obtained with techniques and methods to
be described below, were surprising
\cite{Pirronello1997a,Pirronello1997b}.
From the analysis of our data
three main points stood out.
First, the formation
efficiency obtained from the data was a
steep decaying function of the temperature of the sample
at which the adsorption of
hydrogen atoms took place.
Second,
for sufficiently high surface temperatures,
the efficiency of recombination,
defined here as the probability that two atoms
hitting the surface would recombine,
was lower than predicted by Hollenbach and Salpeter.
Third, it was found that, for
short exposure to atomic
hydrogen so as to guarantee a sparsely populated fraction
of H atoms on the surface
of the dust grain analogue, the formation of molecular
hydrogen obeyed second
order kinetics and that the process of diffusion was
assisted by thermal activation.
The consequence of this finding was
that for a range of high temperatures and low flux,
the molecular hydrogen reaction rate must
depend quadratically rather than linearly on the flux.
Under such conditions the production rate is expressed by

\begin{equation}
R_{\rm H_2} = n_{\rm g} (n_{\rm H} v_{\rm H} \sigma \xi t_{\rm H})^2
                           \alpha/S ,
\label{eq:valerio2}
\end{equation}

\noindent
where $t_{\rm H}$
is the residence time of an atom on the surface and
$\alpha$
is the hopping rate of H atoms between adsorption sites
and $S$ is the number of adsorption sites on a grain.
Here the expression in parenthesis is the coverage
(average number of atoms adsorbed on
the surface).
Biham et al. found that whether the molecular hydrogen formation
rate follows Hollenbach and Salpeter expression or ours depends on whether one is the
situation of fast mobility (and/or high coverage) rather than slow mobility
(and/or low coverage) of H atoms, respectively
\cite{Biham1998}.

The next step was to measure the efficiency on an amorphous carbon
sample
\cite{Pirronello1999},
since amorphous carbon is one of the principal
components of interstellar dust together with silicates
\cite{Mathis1966}.
The efficiency turned out to be higher than in the case of the
polycrystalline sample
and the TPD traces were broader in temperature, suggesting a range of
activation energy barriers for the process of formation and
ejection from the sample.
Considering that the processes
examined here are processes in which the atom/molecule - surface
forces are weak and not very much dependent on the details
of the chemical composition and arrangement of atoms of the solid,
it is reasonable to expect that the differences in
the recombination coefficients
found in the
two types of samples have to be related in a significant part to
the different morphology.
Finally, our group studied the efficiency
and kinetics of molecular hydrogen formation
\cite{Manico2001,Roser2002}
on and the energetics
\cite{Roser2003}
of molecular hydrogen ejection from
the surface of water ice. 
Ice-coated grains are present in dense regions where most
hydrogen is already in molecular form. 
Therefore, the study of the formation of molecular 
hydrogen on icy syrfaces does not have the same urgent interest that the 
formation on refractory materials that are exposed in diffuse clouds.
However, the study of processes of formation of molecular
hydrogen on ice allows the easy manipulation of the surface,
thus offering clues on the processes of hydrogen formation that
in some cases might be extrapolated to other types of dust grains.

In an effort to study how the energy released in the molecular
hydrogen reaction is distributed between the solid and within the
nascent molecule, 
our group 
\cite{Roser2003} and, independently, another group 
\cite{Hornekaer2003}
measured the time of flight of molecules that
were just formed on the surface and were being ejected from it.
The next step, which is under way in
our laboratory and elsewhere
\cite{Perry2002},
is to study the distribution of the energy of the molecule within
the roto-vibrational degrees of freedom.
There are efforts under
way to measure the excitation of the molecule in actual ISM environments;
preliminary studies have not been successful, however, in singling out
excitation of the molecules that can be ascribed to molecule formation
events and not to pumping mechanisms due to shocks and photon absorption
\cite{Takahashi2001,Tine2003}.

\section{Mechanisms of reactions at surfaces}

Knowledge of the reaction steps is important in order to extract values of physical
parameters that can be used then in modeling the chemistry of a cloud.
The standard reaction scheme by which an impinging atom reacts with an atom adsorbed on
a solid surface is the Langmuir-Hinshelwood (LH) mechanism
\cite{Langmuir1918}.
In this reaction, the atom coming from the gas-phase becomes equilibrated with the
surface, diffuses and, if it finds another atom, might react with it.
Broadly speaking, we can identify the following steps in the reaction:
sticking, diffusion, and reaction.
In the sticking, the atom manages to
loose its kinetic energy and becomes thermally accommodated on the surface.
If there is a large mismatch between the mass of the incoming atom and the atom(s)
of the surface, the transfer of energy is inefficient and the atom might bounce a
few times on the surface before it becomes trapped. The energy of binding
of the incoming atom with the surface depends greatly on the type of atoms
and surface and on the way the interaction proceeds.
For example, an atom
approaching a surface will first experience an attractive long-range potential
of the form
$-C_3/z^3$,
where $z$ is the coordinate perpendicular
to the surface
and $C_3$ is a coefficient which depends on the atomic polarizability
and dielectric response of the solid
\cite{Vidali1991}.
As the atom moves along the reaction coordinate it
might get trapped in a shallow (typically $< 500$ K)
potential well (physisorption) or it
might get closer to the surface and, taking advantage of energetically
favorable overlap of electronic orbitals, it might form a stronger bond
($> >500$ K) with atoms on the surface (chemisorption). Whether the incoming
particle will end up in a weak or strong adsorption site might depend on
how the particle approaches the surface, as the particle-surface potential
might have an energy barrier to access a certain adsorption site along the
reaction coordinate.
Indeed, this is the case for H interacting with
the basal plane of graphite. In experiments of interaction of thermal energy
hydrogen atoms with single crystal graphite, it was found that an H atom gets
trapped in shallow sites ($E \sim 32$ meV
\cite{Ghio1980}),
while in other studies 2000 K H
atoms become trapped in strong adsorption sites (chemisorption)
\cite{Zecho2002}.
Theoretical calculations seem to support this picture
\cite{Jeloaica1999,Sha2002},
predicting
that there is a small activation energy barrier for chemisorption.
Because of the high kinetic energy involved,
this type of interaction is not relevant but in very special
astrophysical environments.

Finally, we mention developments in surface science that,
although they emerged in another context, are relevant to this problem.
According to the Eley-Rideal (ER) mechanism
\cite{Harris1981},
the atom from the gas phase impinges on the atom adsorbed on the surface and reacts
with it before becoming equilibrated with the surface.
The two mechanisms, LH and ER,
were not easily distinguished until recently, when advances in surface science
techniques made it possible to measure the energy of the ejected reactant
(expected to be non thermal in the ER case)
or the cross section of the reaction
(a small cross section of atomic dimensions would imply a head-on collision
and therefore the ER mechanism).

A useful way to examine the 
Eley-Rideal mechanism is by
first depositing D atoms and 
then measuring their abstraction by an impinging beam of H atoms,
which leads to the formation of HD molecules.
In experiments reported by Zecho et al. 
\cite{Zecho2002}, D atoms with
2000 K of kinetic energy are deposited on a highly oriented
pyrolitic graphite sample consisting of small (5 $\mu$m in size)
platelets of oriented graphite planes. 
After the deposition, an H
beam is introduced and the rate of the reaction product (HD)
evolving from the surface is measured as a function of time. 
The rate follows an exponential decay as 
$\sim \exp(- \Sigma \Phi_{\rm H} t)$ 
where $\Sigma$ is the cross section for the abstraction
reaction and $\Phi_{\rm H}$ is the flux of H atoms. It is found that
$\Sigma$ is of the order of a few \AA$^2$ indicating that the H atom
reacts directly with the D atom it hits, without prior accommodation
to the surface.

There is also another mechanism, called hot-atom, which is somehow in between the ones
described above.
In this reaction, the atom impinging on the surface maintains some
of its kinetic energy (and/or gain some of the condensation energy) and uses this
energy to sample the surface quickly. The atom suffers multiple collisions with the
surface during the accommodation process and in each collision has a non-zero
probability of reacting with other adsorbed atoms (adatoms).
These mechanisms,
that should be considered as limiting cases of a range of processes leading up
to the reaction, have been observed mostly in the study of the interaction of
hydrogen atoms with well characterized
metal and semiconductor
surfaces.
Although theoretical calculations
\cite{Jackson2001}
suggest that the E-R reactivity
increases as the H-metal bond strength decreases, indicating a preference
for the hot-atom mechanism, it is not clear how those results can be extrapolated
to other types of surfaces.
In Sec. 7 we  discuss the applicability of the different reaction
mechanisms to the formation of H$_2$ on actual grains in the ISM.

\section{Experiments}

\subsection{Experimental methods}

In experiments designed to study processes occurring in the ISM,
one has to mimic the conditions of the particular ISM environment
that one wishes to study. Of the three major requirements for the
study of the formation of molecular hydrogen on dust grain analogues
in the diffuse and dense cloud medium - low (10 - 20 K) sample
temperature, low background pressure (10$^{-10}$ torr),
and low impinging
flux of reactants (hydrogen atoms), the hardest to meet is the third.
As discussed above, the arrival rate of H atoms on a grain surface is
$n_{\rm H} v_{\rm H} \sigma$.
With the parameters used earlier,
one obtains that the flux
per unit area is
$\sim 2.5 \times 10^{6}$
(atoms cm$^{-2}$ sec$^{-1}$).
Unfortunately, this type of
flux is utterly impossible to obtain in laboratory experiments.
For example, the background pressure of 10$^{-10}$ torr already is
equivalent to an arrival rate of $10^{11}$
(atoms cm$^{-2}$ sec$^{-1}$).
If experiments are done with high fluxes one must
have a very fine temporal control of the exposure of the sample
to the incident atoms, so that the exposure (or time during which
the sample is exposed to a flux of impinging particles) is very short.
The reason for this requirement is that ultimately we want to be in
operational conditions of low coverage (that is, fraction of sample
covered by hydrogen atoms); obviously, exposure and coverage are related by
the sticking coefficient. In our experiments we placed great care in reducing
the flux as low as possible. Since the dissociation source that produces atomic
hydrogen from molecular hydrogen can be operated with high (80-90\%)
dissociation
efficiency only in a narrow supply pressure range, in certain experiments,
when we needed to deposit very few atoms, we used a low (5\%)
duty-cycle mechanical chopper.

The experiment is conducted in two steps. In the first one,
the sample is exposed to hydrogen atoms for a given amount of time.
Any molecule generated during this time is detected by a
mass-discriminating detector. According to the scenario
implied by the theory of Hollenbach and Salpeter, molecules
would form at this time since the time it takes an atom to
diffuse and reach another one is orders of magnitude less than the time
during which the sample is irradiated. However, there exists the
possibility that the molecules do not get ejected into the gas
phase after being formed.
Furthermore, it is also possible that
atoms do not have the high mobility envisioned by Hollenbach and Salpeter.
To check for these two latter possibilities, a second experiment is done
in which the sample temperature is rapidly raised and the products evolving
from the surface are collected.
This process is called thermal programmed desorption (TPD).
It can be modeled using the Polanyi-Wigner equation

\begin{equation}
R(t) = \nu N(t)^m \exp(-E_d/k_B T)
\end{equation}

\noindent
where
$R(t)$
is the rate of the atomic/molecular species
departing the surface,
$N(t)$ is the number density of atoms/molecules
on the surface,
$\nu$ is their vibration frequency within adsorption wells
and
$E_d$ is the activation energy for
desorption.
Here, $m$ is the order of the desorption process.
For $m=0$, the desorption rate does not
depend on the number of atoms/molecules on the surface.
This is the case when the surface is covered by
at least several layers of the particles to be desorbed .
For $m=1$ the desorption rate is
linearly proportional to the number density of the adsorbed species.
For $m=2$ the rate depends quadratically on the number density, as in the
formula that we proposed to explain the second order desorption
kinetics we observed for the experiment on the polycrystalline sample.
This is of course a phenomenological equation, and its validity rests
on a number of assumptions
(see Ref. \cite{Menzel1982}),
among which there is the
independence of the desorption energy on coverage
(and this is not too bad an approximation in our case,
since we try to work at very low coverage, down to a few percent of a ML).
The Polanyi-Wigner equation works reasonably well in order to obtain a
preliminary
understanding of the processes being investigated.
However, to confirm the validity
of its application and to obtain more specific information, we used a rate
equation approach (discussed below) to analyze our data.

\begin{figure}
\begin{center}
\includegraphics[width=3in]{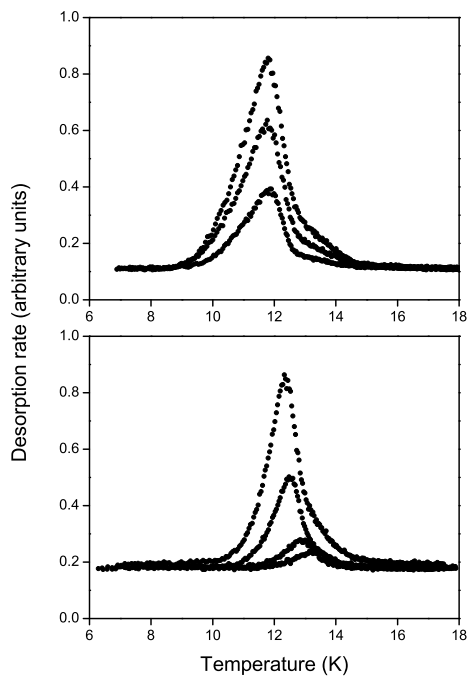}
\end{center}
\caption{
\label{fig:1}
Desorption rate of HD during TPD runs from an olivine slab after
irradiation at surface temperature of 6 K, for long irradiation times
(top) of 2.0, 5.5 and 8.0 minutes and short irradiation times (bottom)
of 0.07, 0.1, 0.25 and 0.55 minutes.
The flux is of the order of 10$^{12}$ atoms/cm$^2$/sec.
See Ref.
\cite{Pirronello1997b}
for details
}
\end{figure}

Fig. 1 shows an example of TPD performed on a polycrystalline olivine sample after the
surface had been irradiated with H and D atoms at the temperature of 7 K
(the reason to use deuterium atoms is explained below).
One can easily note that in the lower panel, corresponding to the shortest
irradiation times, the maximum in the desorption occurs at smaller temperature
as the irradiation time, and consequently the coverage, is increased.
This is a
hallmark of second order desorption kinetics, and the observation of this has
led us to propose
Eq. (\ref{eq:valerio2}).
When the irradiation time is sufficiently long, but
still below one layer of coverage as determined in other experiments, the
position of the maximum does not shift anymore and the shape becomes more
asymmetric, and these are signs of first order desorption.

\subsection{The apparatus}

In order to carry out the measurements described in the previous section,
we have extensively modified an apparatus that was previously used for
surface science experiments of interaction of atomic beams with well
characterized crystalline surfaces. Although the apparatus has been described
in detail previously, here we summarize its main features in the current
configuration.
The apparatus, located in the Physics Department of Syracuse
University (Syracuse, NY- USA) consists of three sections (see Fig. 2):
two atomic/molecular beam lines that can be operated independently, a
sample and detector chamber, and a time-of-flight section. The last two
sections are operated in ultra-high vacuum (UHV) conditions and reach
routinely a background pressure in the $10^{-10}$ torr range.

\begin{figure}
\begin{center}
\includegraphics[width=3.5in]{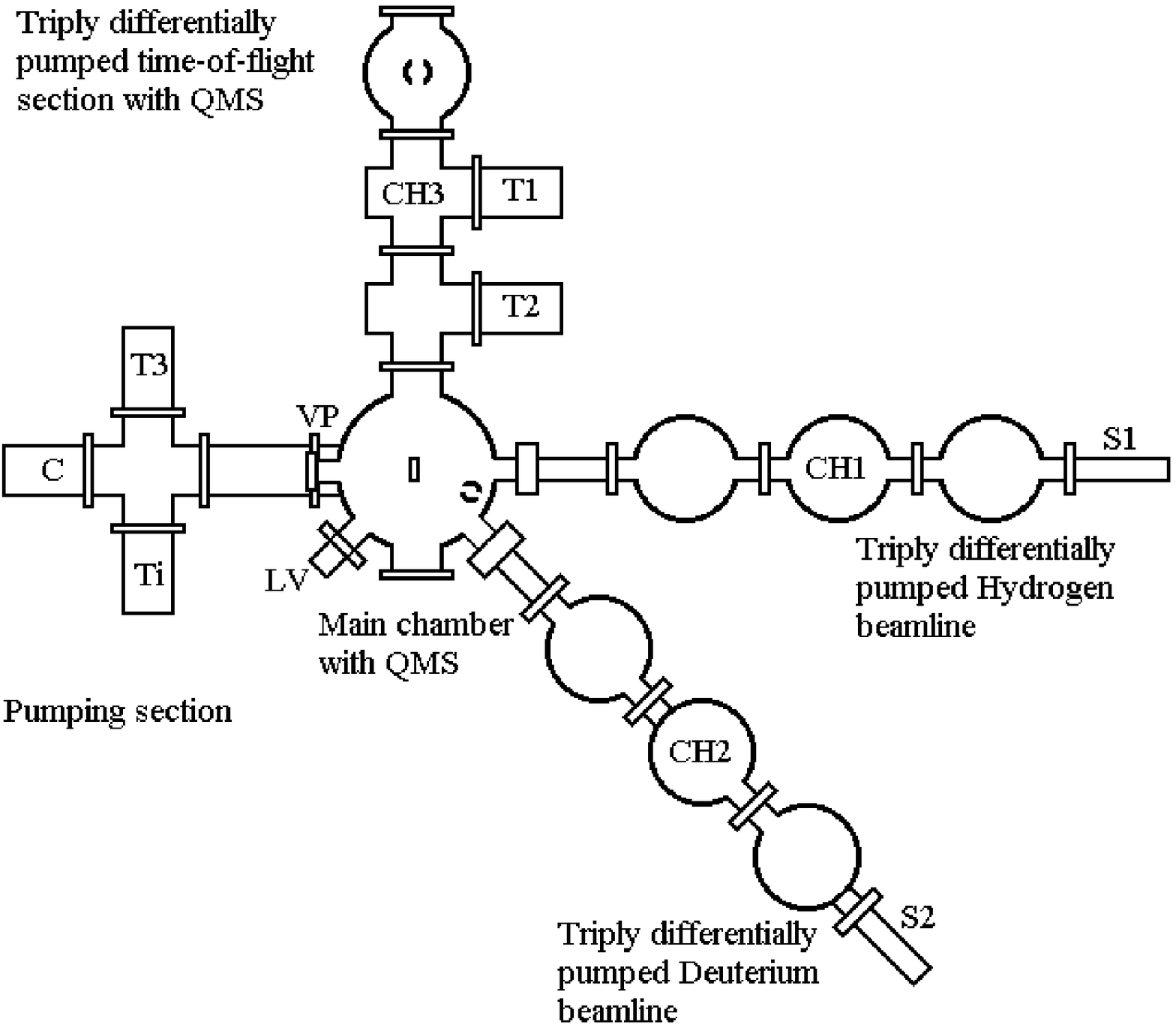}
\end{center}
\caption{
\label{fig:2}
Schematic of the apparatus - top view.
S1 and S2 denote H$_2$ and
D$_2$ radio-frequency dissociation sources; CH1, CH2 and CH3 denote the
positions of mechanical choppers; T1, T2 and T3 are turbopumps while Ti is
a titanium sublimation pump and C a cryopump; LV is the leak valve for
introducing water vapor into the system through a capillary.  See
Ref. \cite{Roser2002} for details.
}
\end{figure}

The reason why two beam lines are used simultaneously, one with
hydrogen and the other with deuterium supply gas, is threefold. First,
even when there is good dissociation, about 10\% of the molecular gas
is not dissociated and therefore is sent to the sample along with the
atomic gas. It would then be difficult to discriminate the formation
of H$_2$ on the surface from this spurious contribution. Second, H$_2$ is
the most abundant residual gas in a well cleaned and baked UHV chamber;
this background gas would become adsorbed on the surface of the sample
and it would be indistinguishable from the formation of H$_2$   occurring
on the sample. Third, if we use H atoms in one line and D atoms in the
other, the only place where they can form HD is on the sample surface.
The product molecule, HD, has mass 3 that is not typically present in a
residual gas, and thus very small amounts of HD coming from the sample
can be detected using a mass spectrometer.

The two beam lines consist each of three differentially pumped sections,
a radiofrequency powered dissociation source and mechanical choppers for
in-phase detection or time-of-flight characterization of the beams.
Each source consists of a Pyrex tube ending with an aluminum nozzle
that can be cooled to 
$\sim$ 
200 K via a copper braid connected to a liquid
nitrogen reservoir. Each source has a water cooled jacket and is
surrounded by a cavity in which the radiofrequency is
coupled to the gas inductively.
 For hydrogen and deuterium dissociation, a total RF
power of about 100 Watts is used.
Dissociation rates obtained are routinely in the 80-90\% as measured using a
mass-discriminating detector (a quadrupole mass spectrometer) located in
the main chamber. The inlet pressure of molecular hydrogen is monitored by
a Baracel or equivalent pressure cell and is regulated using fine
metering valves.

The two beam lines are aimed at a target located in the main UHV chamber.
The sample is mounted on a copper support which is then screwed onto a
Heli-Trans continuous flow cold finger. The temperature of the sample can be
changed by throttling the flow of liquid helium or by using a heater housed in
a ceramic box fastened on the back of the thin vertical copper slab that holds
the sample and a retaining ring.
A silicon diode thermometer and a gold-iron/chromel
thermocouple measure the temperature at the back of the sample.
In a typical run, the
temperature is held fixed between 4.5 and 30 K while the sample is exposed to the
atomic/molecular beams. In a TPD experiment the flow of liquid helium is cut off
and the temperature of the sample rises quickly and in a predictable manner.
To clean the sample during a long set of measurements, the temperature is raised
over 230 K to remove any water deposits by cutting the flow of liquid helium and
using two heaters, the one located behind the sample and another mounted further
up on the cold finger.
To avoid contamination of the sample from background gas
during measurements, a ultra-high vacuum environment is kept by using suitable
materials with low outgassing rates and appropriate pumps. The main chamber is
pumped by a turbopump, an ion pump and a cryopump which assure that when the
sample or the cold finger temperature are suddenly raised for a TPD experiment,
the gases evolving from the sample and its supports are quickly pumped away.
In typical operating conditions, we observed little or negligible back-adsorption
of these gases onto the sample.

The sample can be moved vertically to be positioned in front of the
atomic/molecular beam lines or facing a capillary for water vapor deposition.
The sample holder can also be rotated around its axis in order to face the beam
lines or the time of flight section. To make sure that the atomic/molecular
beams of about 3 mm diameter - hit the same spot on the sample simultaneously,
two He-Ne lasers are placed at the back of the sources when the
apparatus has been
opened up to atmospheric pressure, and the sources and beam
lines are adjusted until
the two laser spots superimpose on the sample.

The detector, a quadrupole mass spectrometer, is housed in a differentially
pumped enclosure with two apertures diametrically located to let the gas
pass through. The detector can be rotated to measure the intensity of each of the
beams; during the measurements of the reaction products it is placed between the
beam lines. In order to measure the dissociation efficiency, the detector
is positioned facing one of the line and is tuned to the molecular mass of
the source gas, i.e. mass 2 for H$_2$ and mass 4 for D$_2$. The change in the detector
signal with the dissociation on and off gives the degree of dissociation; this
method of measurement avoids the problem of recalibrating the efficiency of the
detector for different masses, for example if one wanted to measure the dissociation
tuning the detector first on mass H$_2$  (dissociation off)
and then on H (dissociation on).

The time-of-flight appendix consists of three differentially
pumped chambers separated from each other by 5 mm diameter collimators.
These collimators define a straight-line path from the sample,
through the slots of a constant speed mechanical chopper wheel
mounted in the second vacuum chamber, and through the ionizing
region of a quadrupole mass spectrometer mounted in the third chamber.
The slots of the chopper wheel break the flow of particles desorbing
from the sample into pulses of particles that are admitted into the
free-flight region between chopper wheel and detector.
A LED/photodiode pair is used to detect a triggering slot or
slots on the chopper wheel for synchronizing the detector
with the wheel rotation.

The translational velocity distribution of gas particles within
the pulses can be determined from the known chopper-detector
distance and the flight times of particles across that distance.
In the conventional method of time-of-flight detection, a chopper
wheel with a set of equally spaced, narrow slots is used to admit
pulses of gas into the free-flight region at a regular frequency.
The number of slots and their angular width are chosen for a minimum
of spatial overlap between successive pulses reaching the detector
(with broadening effects such as the dispersion of each pulse over
the free-flight distance taken into account).  This method provides
the advantage that the detector directly records the broadened
time-of-flight distribution of the gas, but with the disadvantage
that the open area of the slots is limited to a few percent of the
area of the chopper wheel.

We have also implemented another method to measure the time of flight.
The cross-correlation method of time-of-flight detection uses a
chopper wheel with a pseudo-random pattern of slots to
intentionally admit spatially overlapping pulses of gas into the
free-flight distance.  The detector integrates these pulses
into an irregular waveform that only reproduces the time-of-flight
distribution in the gas when the waveform is cross-correlated with
the pseudo-random pattern of slots; the pseudo-random
pattern of the slots is chosen for a near optimal transmission of
the time-of-flight distribution through this process.
An advantage of this method of time-of-flight detection over
the conventional method is that the open area of the slots is nearly 50\%,
thus potentially allowing a great reduction in the amount of detection
time required to achieve a given signal-to-noise ratio.
Drawbacks of this method are stringent requirements for the
rotational stability of the chopper motor and the precision
with which the detector is synchronized with the motor,
as well as the additional cross-correlation calculation
needed to extract the time-of-flight distribution from the
detector response.

\section{Results}

For our experiments of H$_2$ formation on dust grain analogues 
we looked at three
classes of materials, each representing a different type of dust grain present
in the ISM: silicates, carbonaceous particles, and ice-coated grains.
For a discussion of different models of dust grains in the ISM, see
\cite{Mathis1966}.

As mentioned before, experiments to measure the recombination efficiency,
defined as the probability that two atoms hitting the surface recombine,
were done in two steps. In the first, while the sample is exposed to beams of
H and D atoms, the HD molecules evolving from the surface are measured.
This efficiency of HD formation turns out to be quite low, at most 10\%,
except at the lowest temperatures ($<10$ K).
Next, after the exposure is completed, the surface
temperature is raised quickly
(the TPD part of the experiment) and HD molecules
coming off the sample are detected,
see Fig. 1 for an example of the rate of desorption as a function of
the temperature of the sample.
Therefore, the trace is proportional to the number of HD
molecules coming off the sample in a temperature interval $dT$.
In order to obtain the
recombination efficiency, the data have to be treated as follows. 
The background
is subtracted out and a correction for the solid angle of the 
detector and the
speed of the molecules going through the detector is applied 
to the value of the
integrated trace. One then has to correct this number because 
some H (or D) atoms
will form H$_2$ (or D$_2$) instead of HD; finally,
the number so obtained is divided by
the measured intensities of the H and D beams
integrated over the time of the exposure.
The experiment is then repeated for exposures at
different sample temperatures, and a
graph as shown in Fig. 3 is obtained.
The exposure used for these experiments is the
shortest that is possible while having a large enough signal/noise ratio
for the analysis.

\begin{figure}
\begin{center}
\includegraphics[width=3in]{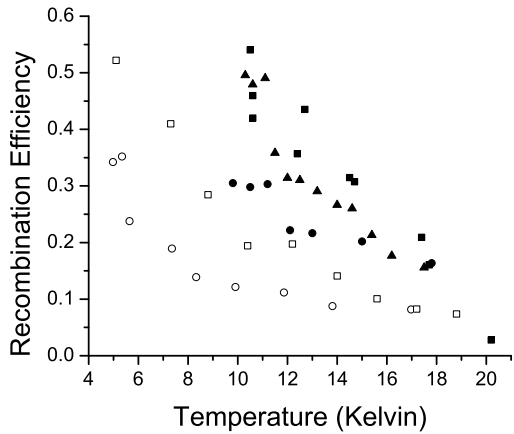}
\end{center}
\caption{ \label{fig:3} Recombination efficiency of molecular
hydrogen vs. sample temperature of H atoms. Filled circles are for
high density amorphous ice, open circles are for low density
amorphous ice prepared by heating high density amorphous ice. Open
squares are for water vapor-deposited low-density amorphous ice. The
scatter in the data points reflects the variability  in the ice
preparation methods. }
\end{figure}

Fig. 4 shows typical traces of TPD for HD desorbing from water ice.
It is known that there are several solid phases of water ice. In the ISM,
the high density amorphous ice phase is the most abundant, but low density
amorphous and crystalline phases have also been detected. Following the recipe
of Jennisken et al.
\cite{Jennisken1995}
who characterized water ice amorphous phases, the
high density phase can be made by depositing water vapor on a cold substrate
at low ($\sim$10 K) temperature. If the ice is warmed past 38 K,
a gradual irreversible
change occurs and the ice transforms in low density ice. 
Finally, at much higher
temperature ($\sim$130 K, depending on the experimental conditions
\cite{Jennisken1994})
low density amorphous ice transforms in crystalline ice
(alternatively, crystalline ice can be obtained by water 
vapor deposition at a
temperature $T > 133$ K
\cite{Petrenko1999}).
We followed the recipe of Jennisken and Blake in order to make
the different types of
amorphous ice, while "crystalline" ice 
was prepared by depositing water vapor at $133 K$.
Details on deposition methods and treatment are given
in Refs.
\cite{Manico2001,Roser2002}.
The important point is that the three different ices give different
TPD traces (because there is no annealing of these samples, traces might
change slightly from
time to time.
Consequently, the shape or relative heights or widths of features in
the TPD traces might change, but overall we observed the 
same type of feature during
the course of our investigations).

\begin{figure}
\begin{center}
\includegraphics[width=3in]{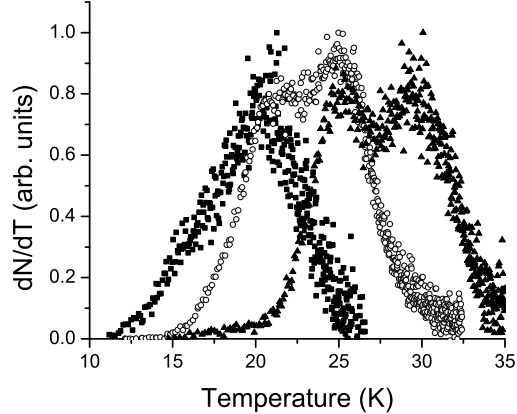}
\end{center}
\caption{
\label{fig:4}
Desorption rate of HD as a function of temperature after
adsorption of H and D at $\sim$10K.
From left to right:  desorption from
crystalline ice, low-density amorphous ice and high-density amorphous ice.
The height of the traces has been normalized for ease of comparison.
See Ref.
\cite{Vidali2004}
for more details.
}
\end{figure}

\begin{figure}
\begin{center}
\includegraphics[width=3in]{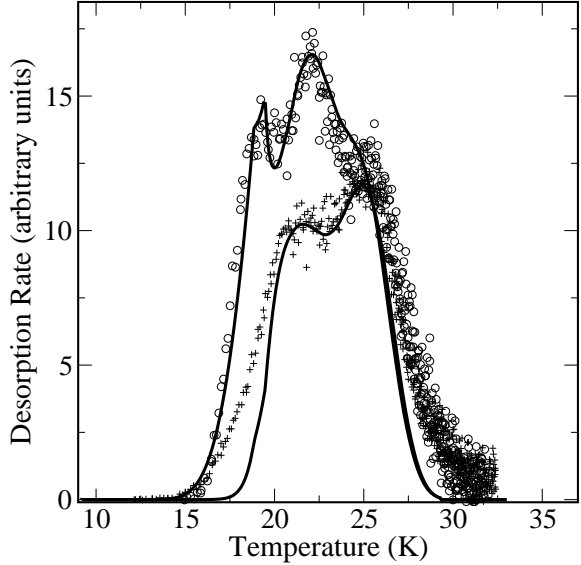}
\end{center}
\caption{
\label{fig:5}
TPD curves of HD desorption after irradiation with
HD molecules ($\circ$) and H+D atoms (+)
on low density ice.
The solid lines are fits obtained by the
rate equations model.
See Ref.
\cite{Perets2005}
for more details.
}
\end{figure}

We mentioned previously that there are different 
possible routes for the 
formation of
molecular hydrogen on the surface of a dust grain.
H$_2$ could form immediately
by the Eley-Rideal process in which an H atom from the gas-phase reacts with
one on the surface. This is unlikely the case here, since the coverage is low
and little H$_2$ is detected coming off the surface 
during the irradiation phase.
Then there is the case of a quick reaction between two 
H atoms that have landed
on the surface and moved rapidly due to diffusion by tunneling. 
In our experiment,
we would expect that the HD molecule just formed on the surface 
would be ejected
out because of the energy released in the reaction, and again we 
measure a very
small signal due to ejection of HD during the irradiation phase. However,
it might be the case that the HD molecule is successful in becoming
thermalized with the surface; in this case it would remain on the surface.
The reason why this could happen is that in a porous solid as 
amorphous 
water ice it
is possible for the nascent molecule to make enough collisions to get
thermalized without getting ejected.
In order to understand
if this is the case and
the mechanisms of reaction, we did an experiment
in which we deposited HD molecules on the surface of ice under the same
conditions as when we deposited H and D atoms. We did then a TPD and
compared the two traces.
If HD formed on the surface and stayed, then
the two traces, one obtained placing HD on the surface and the other
exposing the surface to H and D atoms, would be identical.
But, as one
can see from Fig.
\ref{fig:5},
they are different, indicating that in the experiment
in which the surface is exposed to H and D atoms, the TPD experiment causes
the atoms to migrate, form molecules which are then expelled.
The diffusion and desorption processes of atomic hydrogen on the surface
are controlled by
different energy barriers than the desorption of hydrogen molecules that were
deposited on the surface.
In order to obtain quantitative information,
Perets et al. have used a rate equation model
that we describe in detail below
\cite{Perets2005}.

\begin{center}
\begin{figure}
\begin{center}
\includegraphics[width=3in]{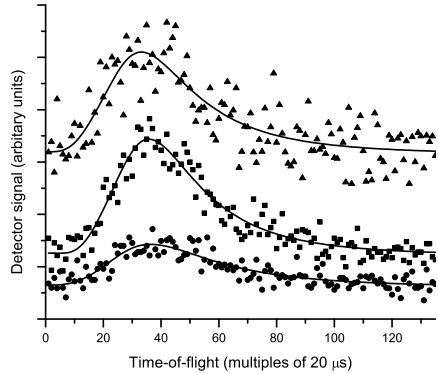}
\end{center}
\caption{
\label{fig:6}
Time-of-flight traces of HD and D$_2$ desorbing from high-density
amorphous ice.  Shown are the sum of two HD spectra taken after 90 minutes
of H and D exposure at a sample temperature of ~8 K (closed triangles), a
D$_2$ spectrum taken after 40 minutes of 
D$_2$ exposure at ~7 K (closed
squares), and a D$_2$ spectrum taken after 40 minutes of D exposure at ~7 K
(closed circles). Solid lines are curve fits with a Maxwell-Boltzmann
function convoluted with a broadening function to correct for instrumental
effects.
See Ref.
\cite{Roser2003}
for more details.
}
\end{figure}
\end{center}

To determine the distribution of kinetic energies of
hydrogen molecules forming and desorbing from the
sample surface, we exposed the sample to H and D atoms
and then ramped the sample temperature for a TPD with the
sample surface rotated to face the time-of-flight line.
In Fig.
\ref{fig:6}
we show a
time-of-flight spectrum of HD
molecules desorbing from a high-density ice layer.
The ice layer was exposed to the H and D beams for
90 minutes at layer temperature of 8 K.
Also shown are time-of-flight
spectra of D$_2$
molecules desorbing from a high-density ice layer.
In one spectrum the sample temperature was 7 K
when it was exposed to the beam of D atoms for 40 minutes.
In the other spectrum the sample was exposed to a beam of D$_2$
molecules.
A 200 Hz chopper motor was used in these three measurements;
Each of the three time-of-flight traces in Fig.
\ref{fig:6}
was fitted
with a distribution of flight times based upon a Maxwell-Boltzmann
distribution of velocities for the particles desorbing from the sample.
For an effusive flow of particles of mass $m$ and temperature $T$ desorbing
from the sample, the number density $n(t)$ of particles with flight times
between $t$ and $t+dt$ is given by

\begin{equation}
n(t) dt \sim f(t) dt \sim {1 \over t^4} \exp
\left({  {-m L^2  } \over {2 k_B T t^2 } } \right),
\end{equation}

\noindent
where $L$
is the chopper to detector flight distance.
Experimentally, each time-of-flight spectrum will also
be broadened due to effects
such as the finite size of the chopper slits or the finite cross-sectional
diameter of the collimating holes
\cite{Verheij1987}.
The function $f(t)$ was
convoluted with a suitable broadening function to take these effects into
account before being used to fit the time-of-flight spectra in
Fig.
\ref{fig:6}.
The kinetic temperature associated with each spectrum can be derived from
the maximum value of $f(t)$:

\begin{equation}
T_{\rm kin} = { {m L^2} \over { 4 k_B t_{\rm peak}^2 } }.
\end{equation}

\noindent
The kinetic temperatures for the time-of-flight spectra in
Fig.
\ref{fig:6} are 24 K for the desorbing HD molecules and
25 K for both spectra of desorbing D$_2$ molecules.
For all three of the time-of-flight spectra in
Fig.
\ref{fig:6},
the kinetic temperature of the desorbing molecules are
roughly similar to the temperature at which the desorption
rate of that molecule is a maximum.
A scenario that can explain 
these results is that a molecule desorbing from
an irregular amorphous ice surface might be more likely to
make one or more collisions with the ice surface.

Hornekaer et al.
\cite{Hornekaer2003,Hornekaer2005}
have used the laser-induced thermal desorption
(LITD) technique to investigate the time-of-flight distribution
of the thermal desorption of HD molecules forming from H and D
exposure of a vacuum-deposited amorphous water ice layer
(which they denote amorphous solid water or ASW).
The LITD technique involves using a laser pulse to
rapidly heat the sample surface on time scales short compared
to those for thermal desorption of the adsorbed species
(see Ref. \cite{Deikhoner2001}
for a detailed description).
Hornekaer et al. reported a peak temperature of 45$\pm$10 K
for the temperature of their ASW ice layer when heated by a laser
pulse and measured time-of-flight distributions for HD and D$_2$
desorbing from the sample consistent with a 45 K Maxwell-Boltzmann
distribution of velocities.  This group's measurements are an independent
determination that the kinetic energy from H and D recombinations on an
amorphous ice surface is lost to the ice layer before the newly formed HD
molecules desorb.

\section{Analysis}

In our TPD experiments most of the adsorbed hydrogen is
released well before a temperature of $30$K is reached.
This indicates that
the hydrogen atoms on the surface are trapped in
physisorption potential wells and are thus only weakly adsorbed.
We also assume that the mechanism for the creation of
H$_{2}$ (or HD) is
the LH scheme, namely that the rate of creation
of H$_{2}$ is diffusion limited
\cite{Pirronello1997a}.

In order to keep the number of parameters at a tractable level,
we do not treat the two populations of H and D adatoms separately.
Instead, we consider only one population of hydrogen atoms
to which we refer as H adatoms.
With this approach, the diffusion, reaction and desorption
processes are controlled by
the activation energy barriers for diffusion and desorption
of H adatoms as well as
the activation energy
barrier for desorption of hydrogen molecules.
In addition, we assume that among the molecules that form on the surface,
some remain adsorbed, while the rest
desorb upon formation.
Such immediate desorption is possible in case that the
4.5 eV of
binding energy of the H$_2$ molecule
is not transformed efficiently into the substrate.

\subsection{Experiments on olivine and amorphous carbon surfaces}

Consider an experiment in which a flux of H atoms
is irradiated on the surface and
a large fraction of them stick.
Once the surface temperature is raised, the adsorbed H atoms
hop like random walkers between adsorption sites on the surface.
When two H atoms
encounter one another they may form an H$_2$ molecule.
Let $N_{\rm H}(t)$ [in monolayers (ML)] be the coverage
of H atoms on the surface and $N_{\rm H_2}(t)$ (also in ML)
the coverage of H$_{2}$ molecules.
We obtain the following set of rate equations:

\begin{subeqnarray}
\label{eq:N}
\dot{N_{\rm H}} & = & F(1 - N_{\rm H} -  N_{\rm H_2})
- W_{\rm H}N_{\rm H} - 2 \alpha N_{\rm H}^{2}
\slabel{eq:N1} \\
\dot{N_{\rm H_2}} & = & \mu \alpha N_{\rm H}^{2} - W_{\rm H_2} N_{\rm H_2}.
\slabel{eq:N2}
\end{subeqnarray}

\noindent
The first term on the right hand side of
Eq.~(\ref{eq:N1})
represents the incoming
flux in the LH kinetics.
In this scheme H atoms deposited on top of H atoms
or H$_{2}$ molecules already on the surface are rejected.
$F$ represents an
{\em effective} flux (in units of ML/sec),
namely it already includes the possibility of a temperature
dependent sticking coefficient.
The second term in
Eq.~(\ref{eq:N1})
represents the desorption of H atoms from the
surface.
The desorption coefficient is

\begin{equation}
W_{\rm H} =  \nu  \exp (- E_{\rm H}^{\rm des} / k_{B} T),
\label{eq:P1}
\end{equation}

\noindent
where $\nu$ is the vibration frequency
(standardly taken to be $10^{12}$ $s^{-1}$),
$E_{\rm H}^{\rm des}$
is the activation energy barrier for desorption
of an H atom and $T$ is the surface temperature.
The third term in
Eq.~(\ref{eq:N1})
accounts for the depletion of the H population
on the surface due to recombination into H$_{2}$ molecules,
where

\begin{equation}
\alpha =  \nu  \exp (- E_{\rm H}^{\rm diff} / k_{B} T)
\label{eq:Alpha}
\end{equation}

\noindent
is the hopping rate of H atoms on the surface
and $E_{\rm H}^{\rm diff}$ is the activation energy barrier for H diffusion.
The first term on the right hand side of
Eq.~(\ref{eq:N2})
represents the creation of H$_{2}$ molecules.
The factor $2$ in the third term of
Eq.~(\ref{eq:N1})
does not appear here since it
takes two H atoms to form one molecule.
The parameter
$0 \le \mu \le 1$
represents the fraction of H$_{2}$ molecules
that remain adsorbed on the surface upon formation,
while the rest
spontaneously desorb due
to the excess energy released in the recombination process.
The second term in
Eq.~(\ref{eq:N2})
describes the desorption of H$_{2}$ molecules.
The desorption coefficient for H$_2$ molecules is

\begin{equation}
W_{\rm H_2}  =  \nu  \exp (- E_{\rm H_2}^{\rm des} / k_{B} T),
\label{eq:P2}
\end{equation}

\noindent
where $E_{\rm H_2}^{\rm des}$ is the activation energy
barrier for H$_{2}$ desorption.
The  production rate $R$ of H$_{2}$ molecules
is given by:

\begin{equation}
R  =  (1-\mu)  \alpha N_{\rm H}^{2} + W_{\rm H_2} N_{\rm H_2}.
\label{eq:Production}
\end{equation}

\noindent
This model can be considered as a generalization of
the Polanyi-Wigner model.
It gives rise to a wider
range of simultaneous applications
and
describes both first order
and second order
desorption kinetics
(or a combination)
for different regimes of temperature and flux.

In the experiments analyzed here,
both the temperature and the flux were controlled
and monitored throughout.
Each experiment consists of two phases.
In the first phase
the sample temperature is constant up to time $t_0$,
under a constant irradiation rate $F_0$.
In the second phase, the irradiation is turned off and
an approximately linear heating of the sample
is applied at the average rate
$b$ (K$/$sec):

\begin{subeqnarray}
\label{eq:temp}
F(t) & = & F_{0}; \ \ \ \ \ \  T(t)  =  T_{0}:
\ \ \ \ \ \ \ \ \ \ \ \ \ \ \ \ \ \ \ 0 \leq t < t_{0}
\slabel{eq:temp1} \\
F(t) & = & 0;    \ \ \ \ \ \ \ T(t)  =  T_{0} + b (t - t_{0}):
\ \ \ \ \  t \geq t_{0}.
\slabel{eq:temp2}
\end{subeqnarray}

\noindent
Here  $T_{0}$ is the constant temperature of
the sample during irradiation.

In order to extract the parameters that are relevant
to the diffusion, reaction and desorption processes of
H atoms on the samples we
performed
numerical integration of
Eq.~(\ref{eq:N}).
The numerically generated
TPD curves were fitted to the experimental ones
by varying the parameters
$E_{\rm H}^{\rm diff}$, $E_{\rm H}^{\rm des}$, $E_{\rm H_2}^{\rm des}$ and $\mu$
until the best fit was obtained.
The parameters that gave rise to the best fits were
$E_0= 24.7$,  $E_1= 32.1$, $E_2= 27.1$ (meV) and $\mu = 0.33$
for olivine,
and
$E_0= 44.0$, $E_1= 56.7$, $E_2= 46.7$ (meV) and $\mu= 0.41$
for amorphous carbon.

Due to the Langmuir rejection mechanism, the coverage of H atoms
on the surface,
$N_{\rm H}(t)$,
does not grow linearly with the irradiation time, $t$,
but is given by

\begin{equation}
N_{\rm H}(t) = 1 - \exp(-F_0 t).
\label{eq:LHrej}
\end{equation}

\noindent
Using this feature we obtained the density of adsorption
sites on the surface.
To this end, the flux densities of the H and D beams were
measured directly.
The total yield  of HD molecules was then fitted to
Eq.~(\ref{eq:LHrej}),
which enabled us to evaluate the flux $F_0$ in ML/sec for each
sample.
We obtained
$F_{0} = 2.7 \cdot 10^{-4}$ (in ML/sec)
for the
olivine experiment and
$F_{0} = 9.87 \cdot 10^{-4}$
for the amorphous carbon experiment.
From these results we found that
the density of adsorption sites on the olivine surface is
$s \cong 2 \cdot 10^{14}$
and for the amorphous carbon surface
$s \cong 5 \cdot 10^{13}$
(sites cm$^{-2}$).

\subsection{Experiments on ice surfaces}

The analysis of the  hydrogen recombination experiments
is complicated because they involve a combination of
diffusion of atoms and molecules,
reaction and desorption processes.
Additional information can be obtained by the analysis
of experiments that involve only molecules,
in which there are no reaction processes and the
results are dominated by molecular desorption.
In these experiments a given amount of molecular hydrogen
is irradiated on the surface, followed by a TPD run.
From the results one can obtain the distribution of energy
barriers for desorption of hydrogen molecules from the surface.
These parameters can then be used in the analysis of
experiments on hydrogen recombination, reducing the number
of fitting parameters used in their analysis.

TPD experiments with irradiation of molecules were recently
performed on amorphous ice surfaces
\cite{Roser2002,Perets2005,Hornekaer2005}.
The TPD curve obtained after irradiation by HD
molecules on low density ice (LDI) is
shown
in Fig.
\ref{fig:5}
(circles).
To fit these TPD curves one needs to assume a
broad distribution of energy barriers for desorption of
hydrogen molecules.
Furthermore,
for low density ice,
the TPD curves are best fitted
by a model that includes three types of adsorption sites for molecules
with different energy barriers for desorption.
Below we present a model that provides a good description of
all the experiments on ice, those with irradiation of atoms
as well as those with irradiation of molecules.
In the model we assume a given density of adsorption sites
on the surface.
Each site can adsorb either an H atom
or an H$_2$ molecule.
In terms of the adsorption of H atoms,
all the adsorption sites are assumed to be identical,
where the energy barrier for H diffusion is
$E_{\rm H}^{\rm diff}$
and the barrier for desorption
is
$E_{\rm H}^{\rm des}$.
As for the adsorption of H$_2$ molecules, we assume that
the adsorption sites may differ from each other.
In particular, we assume that the population of adsorption
sites is divided into
three types.
A fraction $\mu_j$ of the sites belong
to type $j$, where $j=1,\dots,3$,
and
$\sum_j \mu_j = 1$.
The energy barrier for desorption of H$_2$ molecules
from an adsorption site of type $j$ is
$E_{\rm H_2}^{\rm des}(j)$.

Let $N_{\rm H}$
[in monolayers ($ML$)] be the coverage
of H atoms on the surface.
Similarly, let
$N_{\rm H_2}(j)$ (also in $ML$) be
the coverage of H$_{2}$ molecules that are
trapped in adsorption sites of type $j$.
Clearly, this coverage is limited by the number of
sites of type $j$ and therefore
$N_{\rm H_2}(j) \le \mu_j$.
The total coverage of H$_2$ molecules is given by
$N_{\rm H_2} = \sum_{j=1}^{3} {N_{\rm H_2}(j)}$.
Since we assume that each site can host only one atom or one molecule,
the coverage does not exceed a monolayer, and thus
$N_{\rm H} + N_{\rm H_2} \le 1$.
For the case of LDI we thus
obtain the following set of rate equations

\begin{subeqnarray}
\label{eq:Nice}
& \dot{N}_{{\rm H}} & =
F\left(1 - N_{\rm H} - {N_{\rm H_2}} \right) -
W_{\rm H}N_{\rm H} - 2\alpha N_{\rm H}^{2}
\slabel{eq:NH} \\
& \dot{N}_{\rm H_2}(1) &
= \mu_{1} \alpha {N_{\rm H}}^{2} - W_{\rm H_2}(1)N_{\rm H_2}(1)
\slabel{eq:N2(1)} \\
& \dot{N}_{\rm H_2}(2) & =
\mu_{2} \alpha {N_{\rm H}}^{2} - W_{\rm H_2}(2)N_{\rm H_2}(2)
\slabel{eq:N2(2)}\\
& \dot{N}_{\rm H_2}(3) & =
\mu_{3} \alpha {N_{\rm H}}^{2} - W_{\rm H_2}(3)N_{\rm H_2}(3).
\slabel{eq:N2(3)}
\end{subeqnarray}

\noindent
Eq.
(\ref{eq:NH})
is identical to
Eq.
(\ref{eq:N1}).
Eqs.~(\ref{eq:N2(1)})-(\ref{eq:N2(3)})
describe the population of molecules on the surface.
The first term on the right hand side of
each of these three equations
represents the formation of H$_2$ molecules
that become adsorbed in a site of type
$j=1$, 2 or 3.
The second term in
Eqs.~(\ref{eq:N2(1)})-(\ref{eq:N2(3)})
describes the desorption of H$_2$ molecules from
sites of type $j$,
where

\begin{equation}
W_{\rm H_2}(j)  =
\nu \exp (- E_{\rm H_2}^{\rm des}(j) / k_{B} T)
\label{eq:Wk}
\end{equation}

\noindent
is the H$_2$ desorption coefficient and
$E_{\rm H_2}^{\rm{des}}(j)$
is the activation energy
barrier for H$_2$ desorption from an adsorption site
of type $j$.
The H$_2$ production rate $R$ is given by:

\begin{equation}
R  =  \sum_{j=1}^{3}{W_{\rm H_2}(j) N_{\rm H_2}(j)}.
\label{eq:Production2}
\end{equation}

\noindent
The main difference between this model and the model
used previously in the analysis of the olivine and carbon experiments
is in the way molecules desorb from the surface.
In the earlier model, a fraction $1-\mu$ of the molecules,
desorb from the surface upon formation.
This was interpreted as a result of
the 4.5 eV of binding energy which is not
transferred efficiently into internal
degrees of freedom of the grain.
In the model used for the ice experiments,
hydrogen molecules do not desorb upon formation but
become trapped in adsorption sites.
This requires an efficient transfer of the the bidning energy
into the grain. On porous surfaces, repeated collisions of the
molecule with the surface are possible, which may help to transfer
the energy to the grain.

The model used for the analysis of the ice experiments is motivated
by the results of the experiments with irradiation by molecules
shown in Fig.
\ref{fig:5}.
The TPD curve of HD desorption obtained after irradiation with
H and D atoms
is also shown in Fig.
\ref{fig:5}.
A comparison between the TPD curves obtained after irradiation
by HD molecules
and those obtained after irradiation by H and D atoms on LDI
provides strong evidence for the proposed model.
In experiments with irradiation of hydrogen atoms,
the TPD curve consists of two peaks.
These peaks coincide with the two higher peaks, among the
three that are obtained after irradiation with molecules.
The lowest peak is wiped out.
This is because in the experiments in which atoms are irradiated,
molecules are formed only at a later stage, after the temperature
has been raised.
At that time, the binding energy at the
sites that correspond to the lowest peak is already insufficient
for binding hydrogen molecules.

\subsection{Astrophysical implications}

Although the experiments that are described below were done with a very
low flux of hydrogen atoms, the laboratory flux is still orders of
magnitude higher than the one impinging on an actual dust grain in the ISM.
Therefore, it is necessary to use theoretical tools in order to extract
from experimental data information pertinent to the ISM.
We do the following.
First, once a preliminary analysis of the data has individuated the main
physical processes at play, the theoretical model is built and the parameters
are obtained from a fit to the data.
Then, once it has been verified that the
model is viable, the theory, using the parameters coming from the experiments, can
be used to predict the results of processes that are actually happening in the ISM.
Katz et al. used rate equations to fit the data of hydrogen
recombination on amorphous carbon and polycrystalline olivine
\cite{Katz1999}.
The activation energy barriers for the relevant diffusion and desorption processes
were extracted from the data.
These barriers were used in order to calculate
the efficiency of recombination for a range of
low fluxes of hydrogen atoms and grain temperatures which are
typical in the interstellar space
\cite{Katz1999}.

** In the paragraph below you present the data from the Perets et
al. paper - since in the paragraph above the work of Katz et al. is
mentioned, it is would be better to specify clearly that the work
mentioned below is done using  the model presented in Section 6.2 **

\begin{center}
\begin{figure}
\begin{center}
\includegraphics[width=3in]{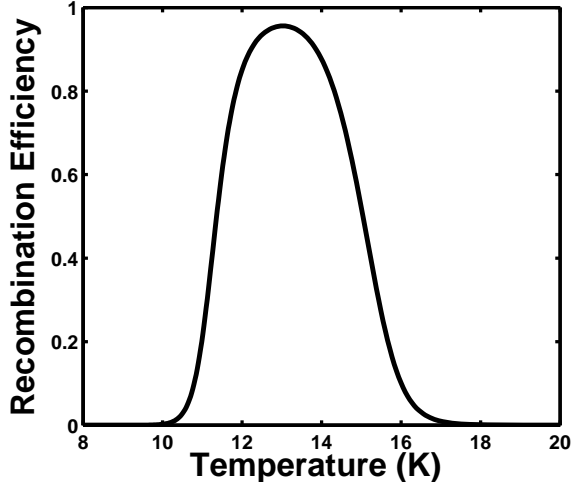}
\end{center}
\caption{
\label{fig:7}
Recombination efficiency of molecular hydrogen formation on LDI vs.
surface temperature.
High efficiency is obtained in the temperature range between 11-16 K.
See Ref.
\cite{Perets2005}
for more details.
}
\end{figure}
\end{center}

The experimental results indicate that the mobility of H atoms on
the surface is dominated by thermal hopping rather than tunneling.
To examine the astrophysical implications of the results we used the
rate equations to calculate the formation rate of H$_2$ under
conditions that are typical in diffuse clouds. In Fig. \ref{fig:7}
we present the recombination efficiency of H$_2$ molecules vs.
surface temperature for LDI with $5 \times 10^{13}$ adsorption sites
per cm$^{2}$, exposed to a flux of H atoms from the gas phase, with
density of $n_{\rm H}=10$ (cm$^{-3}$) and gas temperature ofn 100 K.
This calculation, for the ice surface was done using Eq.
(\ref{eq:Nice}).

It turns out that
the temperature range in which H atoms are
highly mobile on the surface but resides on it long enough
to find each other and recombine is rather narrow.
For ice, it is between 11 K and 16 K.
At higher temperatures atoms desorb from the surface before
they have sufficient time to encounter each other.
At lower temperatures diffusion is suppressed while
the Langmuir rejection leads to saturation of the surface
with immobile H atoms and recombination is suppressed.

In general,
the temperature range in which H$_2$ formation
is highly efficient
is given by

\begin{equation}
\frac{E_H^{\rm diff}}{k_{\rm B} \ln(\nu /F)} < T <
\frac{2 E_H^{\rm des} - E_H^{\rm diff}}{k_{\rm B} \ln(\nu /F)}.
\end{equation}

\noindent The width of this range is thus proportional to the
difference between the diffusion and desorption barriers of H atoms
on the surface. For the 
olivine sample, under typical flux that
exists in diffuse clouds, the temperature window of high efficiency
was found to be in the range $8  < T < 12$ (K), while for amorphous
carbon it is $12  < T < 16$ (K).

\section{Discussion}

\subsection{Molecular hydrogen formation in photon-dominated regions}

The results of the laboratory experiments indicate that a combination
silicate and carbon grains may serve as efficient catalysts for
H$_2$ formation in diffuse and dense molecular clouds. 
Given the fact that there must be some mobility of H atoms on surfaces of grains in the 
temperature range 10 to 20 K, the binding of H 
on grains must be governed by weak  physisorption 
forces. 
However, molecular hydrogen has also been observed in photon-dominated
regions (PDR's), where the temperature of dust grains is typically
between 30-50 K
\cite{Hollenbach1997,Habart2004}.
These high temperatures are
caused by  
far untraviolet photons, which penetrate into the interstellar clouds
and heat the gas and dust in these regions.
In particular, dust grains are heated
through photoelectric heating
while H$_2$ molecules are heated through photon 
absorption.
The surface recombination processes
observed in the laboratory experiments cannot produce
molecular hydrogen at these high surface temperatures.
At these temperatures,
the residence time of hydrogen atoms
in physisorption sites
on the surface is too short for recombination to take place.

In order to explain the existence of molecular hydrogen in PDR's,
Hollenbach and Salpeter \cite{Hollenbach1971a}
suggested that the grain surfaces contain enhanced
adsorption sites with ``semi-chemical binding'',
where atoms can stick much stronger.
As a result,
these atoms stay longer on the surface, allowing more time
for pairs of hydrogen atoms to
recombine before they desorb.
For carbon materials,
it was shown that these enhanced
binding sites can be
chemisorption sites, and
their binding energies were calculated
\cite{Bennet1971,Dovesi1976,Dovesi1981,Aronowitz1985}.

According to recent analysis, the chemisorption model predicts high
recombination efficiency 
at PDR temperatures, given that the energy
barriers for entering the chemisorption sites are not larger than
the binding energies of the physisorption sites, 
i.e., $\sim$ 
50 meV \cite{Cazaux2004,Habart2004}. However, ab initio
calculations show that the energy barrier for an H atom to enter a
chemisorption site on a graphite surface is $\simeq$ 0.2 eV
\cite{Jeloaica1999,Sha2002a,Sha2002b}. 
Such barriers prevent H atoms
from entering chemisorption sites at grain temperatures lower than
100 K. These results were confirmed by recent experiments in which
an activation energy barrier of 0.18 eV was found for hydrogen on
C(0001), \cite{Zecho2004}. 
If such barriers exist also on the amorphous surfaces of interstellar dust
grains, chemisorption sites
are not likely to play a major role in H$_2$ formation on dust
grains in PDR's.

\subsection{The role of chemisorption sites}

The experiments of H$_2$ formation on amorphous carbon and polished
olivine have also been analysed  by Cazaux and Tielens
\cite{Cazaux2002,Cazaux2004} using a different, more complex model.
They introduced additional free parameters with respect to the Katz
et. al. model and also extended their analysis to the formation of
H$_2$ at much higher temperatures. The differences between this
model and the Katz et. al. model are as follows. First, Cazaux and
Tielens take into account the presence of both physisorbed {\it and}
chemisorbed sites on the surface. Second, their model allows quantum
mechanical diffusion in addition to the thermal hopping of absorbed
H atoms. Third, they treat separately the H and D isotopes used in
the experiments. Using this model Cazaux and Tielens fitted the
experimental TPD curves obtained in the polycrystalline olivine and
amorphous carbon experiments. They found that quantum mechanical
tunneling between physisorption sites is too slow to be
important, supporting the Katz et. al. model. Their analysis
suggested that it may be important (under certain assumptions on the
energy barriers between 
physisorption
and chemisorption sites) to
populate chemisorption sites at low temperatures. Results similar to
the ones in Katz et. al. were obtained for the energy barriers of
the physisorbed sites. Additional experiments at much higher
temperatures 
on a graphite surface
\cite{Zecho2002} were used in order to obtain
information on the energy barriers for the chemisorption sites.
However, the results of the high temperature experiments are not
fitted well using this model. This may indicate the possibility that
additional processes take place under these conditions.

Other parameters of the Cazaux and Tielens'
model have been only partially constrained by
the experimental data.
Specifically, Cazaux and Tielens showed
that chemisorption
sites could have an important role in H$_2$ formation at low
temperatures
only if the energy barrier for entering a chemisorption site is much
lower than expected by theory and found in recent experiments
\cite{Jeloaica1999,Sha2002,Zecho2002}.

Thus, the chemisorption and tunneling processes suggested in this model do not
seem to play an important role in the current experiments and in the
conditions relevant for diffuse interstellar clouds and in PDRs.
Differences in
the behavior of H and D isotopes and their consequences that are
considered in the
model may be important, but need experimental evidence, which cannot
be extracted from current experiments where measurements of H$_2$ and
D$_2$ production have not been done.

\subsection{Recent experiments by Hornekaer et al. on ice surfaces}

Recently, Hornekaer et al.
\cite{Hornekaer2003}
presented interesting results
on H$+$D recombination on porous and non-porous amorphous solid water (ASW).
Amorphous ice
is considered a good
analog for ice mantles on grains in dark clouds.
Their experiments were performed under conditions
and using an equipment not
too different from the one used by our group.
Their atom beam fluxes were
$\sim$ $10^{13}$
(atoms cm$^{-2}$ sec$^{-1}$, namely
about one order of magnitude higher than the
fluxes used in our experiments.
The range of exposure times was comparable.
Hornekaer et al.
investigated the kinetics of HD formation and measured
the efficiency of recombination and the energetics of
the molecules released from the ice layer after formation.
The efficiency values they obtained are close to those obtained by our group
on amorphous ice.
The energy distribution of molecules formed showed
that at least in porous amorphous ice, molecules are
thermalized by collisions with the walls of pores (where they are
formed)
before they emerge into the gas phase.

Hornekaer et al. performed
TPD experiments in which they irradiated H and D
atoms either simultaneously or sequentially
after waiting a delay time interval before
dosing the other isotope.
On porous ASW the results they obtained are
consistent with a recombination occurring quickly after atom dosing
due to a high mobility of the adsorbed atoms
even at temperature as low as 8 K.
This high mobility was attributed either to quantum
mechanical diffusion or to the so called hot atom mechanism,
where thermal activation is not likely to play a significant
role at this temperature.
This conclusion is sensible in light of the
high coverages of H and D atoms irradiated in these experiments,
which required the adsorbed atoms to diffuse only short distances before
encountering each other.
The hot atom mechanism may be able to provide the required mobility.
In this case H and D atoms retain a good
fraction of their gas phase kinetic
energy during the accommodation process
\cite{Buch1991,Takahashi2001}.
This enables them
to travel on the ice surface and inside its pores
for several tens of Angstroms exploring several adsorption sites and
recombining upon encountering already adsorbed atoms.

The reconciliation of their results with ours, which were obtained
with low fluxes of atoms and short irradiation times, and hence at
low coverages, comes from the fact that in our conditions the number
of sites explored  (which should be the same as in Hornekaer et
al.'s experiment) by the hot atom during accommodation is not
sufficient for this atom to encounter an already adsorbed atom and
react with it with a significant probability. 
Indeed, this is
confirmed quantitatively by a preliminary analysis (to be published
in a forthcoming paper) of the irradiation process,
using rate equations, in which the probability of reaction to form H$_2$ is
proportional to the region of surface spanned by the hot atom.
The solution of these equations, i.e. the number of
H$_2$ molecules produced as a function of time during the
adsorption phase, shows that under Hornekaer et al.'s conditions,
H$_2$ molecules are readily formed, while in ours they are not.
Therefore, in Hornekaer et al.'s case, all of the H$_2$ is formed by
hot atoms making a few hops and encountering other previously
adsorbed atoms. 
As observed experimentally by Hornekaer et al., the
molecules remain trapped in the ice until the TPD is initiated. In
our experiment, the atoms are too far apart from each other during
the deposition phase, and the theoretical analysis confirms our
original interpretation 
\cite{Roser2002}
for H$_2$ formation on ice,
and 
\cite{Pirronello1997a,Pirronello1997b,Katz1999} 
for H$_2$
formation on olivine and amorphous carbon that atoms remain
confined to their sites until the TPD is initiated.

\subsection{The effect of grain size on H$_2$ formation efficiency}

Rate equations are an ideal tool for the simulation of
surface reactions, due to their simplicity and high computational efficiency.
In particular, they account correctly for the temperature dependence
of the reaction rates.
However, the formation of molecular hydrogen in the interstellar medium
takes place on grains of sub-micron size, under extremely low flux.
In this case rate equations may not be suitable
because they ignore the fluctuations as
well as the discrete nature of the population of H atoms on each grain
\cite{Charnley1997,Caselli1998,Shalabiea1998,Stantcheva2001}.
For example, as the number of H atoms on a grain fluctuates
in the range of 0, 1 or 2,
the H$_2$ formation rate cannot be obtained from the average
number alone.
Recently, a master equation approach was proposed, that
takes into account both the discrete nature of the
population of H atoms as well as the fluctuations, and
is thus suitable for the simulation of H$_2$ formation on
interstellar dust grains
\cite{Biham2001,Green2001}.
Since interstellar dust grains exhibits a broad distribution of sizes
covering much of the range between a micron and a nano-meter, it is
important to take the effect of grain size into account.
However, the parameters obtained from the experiments, using rate
equation analysis remain valid.
Inserting these parameters into the master equation is expected to provide
reliable results for the production rate of molecular hydrogen in interstellar
clouds.

\section{Summary}

For more than a quarter of a century, the  model of 
Hollenbach and Salpeter has stood as the 
reference point for the description of the molecular 
hydrogen formation on dust grain surfaces. 
Our experiments, the results of which were first published in 1997
\cite{Pirronello1997a, Pirronello1997b}, 
showed that for certain models of grains the efficiency of 
recombination was below expectations; that a high recombination 
efficiency was obtained only within a narrow range of temperature of the grain; 
and that the kinetics of reaction hinted to hitherto neglected processes 
of mobility of H atoms on surfaces. Subsequent experiments on amorphous 
surfaces (amorphous carbon grains and two different phases of amorphous water ice) 
gave higher efficiency of recombination and over a wider temperature range, 
suggesting the possibility that morphology plays a substantial part. 
It also became clear that weak physisorption forces were responsible 
for the interaction of H atoms with the surface.  In the meantime, 
theoretical work showed that the Hollenbach and Salpeter model and 
our model could both be responsible for the molecular hydrogen production, 
but under different ISM conditions
\cite{Biham1998}. 
In the last couple of years, other experimental and theoretical 
groups became interested in this problem. 
The work of Hornekaer et al. 
\cite{Hornekaer2003}
confirmed our findings on the formation efficiency of H$_2$ on amorphous 
water ice
and on the energetics of ejection, although their interpretation 
of their results (fast formation of molecules via athermal mechanisms) 
seemed at first sight
to be at odds with  
our analysis of our data
(H diffusion is initiated by thermal activation). 
We have shown in this paper that the different interpretations stem from the 
fact that the two experiments were done with fluxes differing by more than one 
order of magnitude. 
Specifically, the higher flux in the Hornekaer et al. 
experiments may have caused certain mechanisms, 
such as the hot atom mechanism, to 
become prominent and yield the rapid formation of molecules,
while in our case the leading initiator of reactions is the
thermally activated mobility. 
The theoretical work of Cazaux and Tielens 
\cite{Cazaux2004}
extended existing models to the 
case in which chemisorption sites are present on the surface, while confirming 
that at the low temperatures relevant to grains in quiescent 
diffuse and dense clouds, 
the laboratory  results could be explained using physisorption forces and 
thermally activated diffusion. 
Recent progress in probing the mechanisms of formation of 
molecules on surfaces at higher 
temperatures 
\cite{Zecho2002} further improve the 
understanding of H$_2$ formation in a wider range of ISM conditions.

\section{Acknowledgments}

This work was supported
by NASA through grants NAG5-6822 and NAG5-9093 (G.V),
by the Italian Ministry for University and Scientific Research
through grant 21043088 (V.P)
and by
The Adler Foundation for Space Research
and the Israel Science Foundation (O.B).

\end{document}